\documentclass[prl,superscriptaddress,preprint,endfloats,showpacs]{revtex4}

\usepackage{amsmath, amsthm, amssymb, pifont, wasysym}

\usepackage{graphicx}
\usepackage{bm} 
\usepackage{pifont}
\usepackage{dcolumn} 
\usepackage{color}

\newcommand {\SRO}{Sr$_2$RuO$_4$}
\newcommand {\Tc}{$T_{\mathrm{c}}$}
\newcommand {\Hctwo}{$H_{\mathrm{c}2}$}
\newcommand {\Hparaa}{$H \parallel a$}
\newcommand {\Hparac}{$H \parallel c$}

\newcommand {\Hctwoa}{$H_{\mathrm{c}2\parallel a}$}
\newcommand {\Hctwoc}{$H_{\mathrm{c}2\parallel c}$}
\newcommand {\pxy}{$p_{x}\pm ip_{y}$}
\begin{document}

\title{Anomalous Enhancement of Upper Critical Field\\
in Sr$_{2}$RuO$_{4}$ Thin Films}

\author{Masaki Uchida}
\email[Author to whom correspondence should be addressed: ]{uchida@ap.t.u-tokyo.ac.jp}
\affiliation{Department of Applied Physics and Quantum-Phase Electronics Center (QPEC), University of Tokyo, Tokyo 113-8656, Japan}
\affiliation{PRESTO, Japan Science and Technology Agency (JST), Tokyo 102-0076, Japan}
\author{Motoharu Ide}
\affiliation{Department of Applied Physics and Quantum-Phase Electronics Center (QPEC), University of Tokyo, Tokyo 113-8656, Japan}
\author{Minoru Kawamura}
\affiliation{RIKEN Center for Emergent Matter Science (CEMS), Wako 351-0198, Japan}
\author{Kei S. Takahashi}
\affiliation{PRESTO, Japan Science and Technology Agency (JST), Tokyo 102-0076, Japan}
\affiliation{RIKEN Center for Emergent Matter Science (CEMS), Wako 351-0198, Japan}
\author{Yusuke Kozuka}
\affiliation{Department of Applied Physics and Quantum-Phase Electronics Center (QPEC), University of Tokyo, Tokyo 113-8656, Japan}
\author{Yoshinori Tokura}
\affiliation{Department of Applied Physics and Quantum-Phase Electronics Center (QPEC), University of Tokyo, Tokyo 113-8656, Japan}
\affiliation{RIKEN Center for Emergent Matter Science (CEMS), Wako 351-0198, Japan}
\author{Masashi Kawasaki}
\affiliation{Department of Applied Physics and Quantum-Phase Electronics Center (QPEC), University of Tokyo, Tokyo 113-8656, Japan}
\affiliation{RIKEN Center for Emergent Matter Science (CEMS), Wako 351-0198, Japan}

\begin{abstract}
We report large enhancement of upper critical field {\Hctwo} observed in superconducting {\SRO} thin films. Through dimensional crossover approaching two dimensions,  {\Hctwo} except the in-plane field direction is dramatically enhanced compared to bulks, following a definite relation distinct from bulk one between {\Hctwo} and the transition temperature. The anomalous enhancement of {\Hctwo} is highly suggestive of important changes of the superconducting properties, possibly accompanied with rotation of the triplet $d$-vector. Our findings will become a crucial step to further explore exotic properties by employing {\SRO} thin films.
\end{abstract}
\maketitle

Superconductors with a multicomponent order parameter, represented by spin-triplet superconductors, have attracted great interest as a ground of rich physics originating in the internal degrees of freedom. Among them, a layered-perovskite superconductor {\SRO} has been a leading candidate possibly having chiral $p$-wave symmetry \cite{SRO, SROreview1, SROreview2}, which is one of topological superconducting states supporting Majorana modes at edges and vortices \cite{SROreview3desiringfilm, SROreview4desiringfilm}. For further investigation and possible applications of the unique properties, the use of {\SRO} thin films has been increasingly demanded in recent years \cite{SROreview3desiringfilm, SROreview4desiringfilm}.

In general, bulk superconducting state or pairing symmetry can be altered in the thin film form, affected by dimensionality change, inversion symmetry breaking, and/or epitaxial strain \cite{book}. Spin-triplet superconducting states are characterized by $d$-vector, which represents the pair amplitude for the spin component perpendicular to the corresponding basis. Particularly in the case of {\SRO}, it has been theoretically suggested that the $d$-vector can flip from perpendicular (chiral $p$-wave) to parallel to the RuO$_2$ $ab$ plane in the reduced dimensions, while the system still can host the Majorana modes \cite{thinSROdvector}. Also, in helium-3 superfluid phases, changes of the $p$-wave order parameter have been experimentally demonstrated by mesoscopically confining it in a two-dimensional (2D) cavity \cite{thin3Hedvector}. In this context, it is indispensable to examine fundamental superconducting properties of {\SRO} thin films. While growth of the superconducting films had been extremely challenging over the past decades since the discovery of {\SRO} \cite{YoshiharuPLD, RobinsonPLD}, the reproducible and controllable growth has been recently achieved by refining molecular beam epitaxy techniques \cite{UchidaMBE, HariMBE}.

Upper critical field {\Hctwo} is one of the fundamental superconducting parameters related to superconducting symmetry, and thus has been intensively investigated in the study of {\SRO} bulks \cite{Hc2kittaka_graph, Hc2akimacohelength_graph, Hc2mackenzie_graph, Hc2yoshida_graph, Hc2deguchigap2, Hc2deguchigap1phase2, Hc2Pauli1, Hc2Pauli2, Hc2yonezawafirst1, Hc2eutecticRu_3K1, Hc2eutecticRu_3K4, Hc2inplane, Hc2calc2inplane, Hc2strain, WHHforSRO1}. While its behavior is generally consistent, some features have been interpreted as incompatible with the simple {\pxy} model \cite{SROreview4desiringfilm}. In particular, {\Hctwo} observed for the in-plane field direction is much more suppressed than expected at low temperatures, also accompanied with the first-order superconducting transition \cite{Hc2kittaka_graph, Hc2yonezawafirst1}. This suppression implies that {\Hctwo} for {\Hparaa} might be affected by the paramagnetic pair breaking induced by the Zeeman splitting, called the Pauli limit \cite{SROreview4desiringfilm}.

Here we report detailed dependences of {\Hctwo} in {\SRO} thin films, by measuring low-temperature magnetotransport systematically changing the field angle. The superconducting films are grown on a lattice-matched cubic substrate, yielding extremely limited defects in the films \cite{UchidaMBE}. In addition to dimensional crossover confirmed in the field angle and temperature dependences, {\Hctwo} in the films is largely enhanced over a wide range of field angles except the in-plane direction, up to about four times the bulk value. This anomalous enhancement indicates that the triplet $d$-vector in thin films may be aligned on the $ab$ plane, consistent with the recent theoretical prediction \cite{thinSROdvector}.

Superconducting single crystalline {\SRO} films as displayed in Figs. 1(a) and (b) were epitaxially grown on cubic (LaAlO$_{3}$)$_{0.3}$(SrAl$_{0.5}$Ta$_{0.5}$O$_{3}$)$_{0.7}$ (LSAT) (001) substrates by oxide molecular beam epitaxy, following the same procedures detailed in Ref. \cite{UchidaMBE}. Sr and Ru elemental fluxes were simultaneously supplied from a conventional Knudsen cell and an electron beam evaporator, respectively. The deposition was performed flowing distilled 100\% ozone with a pressure of $1\times 10^{-6}$ Torr and heating the substrate at 900 $^{\circ}$C. The film thickness is typically 50 nm along the $c$ axis and the channel area of each sample is approximately 500 $\mu$m $\times$ 200 $\mu$m in the $ab$ plane. Four-point measurements of the longitudinal resistivity were performed using low-frequency lock-in techniques with an excitation current of 3 $\mu$A along the $a$ axis. Two samples were cooled down to 60 mK in a $^3$He-$^4$He dilution refrigerator equipped with a superconducting magnet. As shown in Figs. 1(c) and (d), a superconducting transition with $T_{\mathrm{c}} \sim 1.3$ K (onset) is confirmed for a typical sample. While the present films do not yet reach the high standard quality of {\SRO} bulk single crystals \cite{highestbulk}, the transition temperature and its sharpness are now qualitatively comparable to the first reported bulk single crystal \cite{SRO}. For field rotation in the $ac$ plane, the samples were set on a single-axis rotating stage mounted on the mixing chamber.

Figures 1(e) and (f) show field dependence of the in-plane resistivity in a 50 nm thick {\SRO} film, taken for {\Hparaa} and {\Hparac} geometries at various temperatures down to 60 mK. Unlike the Ru eutectic phase \cite{Hc2eutecticRu_3K1, Hc2eutecticRu_3K4} or uniaxially strained phase \cite{Hc2strain}, hysteresis between the upward and downward sweeps is not detected in the resistivity. With increasing field, the resistivity changes from zero to a normal-state value due to the suppression of superconductivity through {\Hctwo}. Reflecting anisotropic superconductivity of this compound, the superconducting state is maintained up to higher fields for {\Hparaa} than for {\Hparac}.

Detailed field angle dependence of {\Hctwo} approaches 2D behavior with reducing the system thickness. Figure 2 compares field angle dependence between a {\SRO} bulk \cite{Hc2kittaka_graph} and the film, where the out-of-plane field angle $\theta$ is measured from the $a$ axis. In the bulk, the angle dependence except for a very low angle region is well described by the following  anisotropic 3D Ginzburg-Landau (GL) model \cite{Hc2kittaka_graph, Hc2yonezawafirst1}.
\begin{eqnarray}
\left(\frac{H_{\mathrm{c2}}(\theta)\sin \theta}{H_{\mathrm{c2}||c}}\right)^2+\left(\frac{H_{\mathrm{c2}}(\theta)\cos \theta}{H_{\mathrm{c2}||a}}\right)^2=1
\end{eqnarray}
{\Hctwo} is assumed to be dominated by the diamagnetic pair breaking process originating from the screening currents, known as the orbital limit. The coherence length along the $c$ axis $\xi_{c}$, calculated from the GL expression $\xi_{c}=\sqrt{\Phi_{0}H_{\mathrm{c2}||c}/2\pi {H_{\mathrm{c2}||a}}^2}$, is 3.2 nm, which is much larger than the lattice spacing of the RuO$_2$ layers. In this regard, superconductivity in the {\SRO} bulk is not classified into ideal 2D systems \cite{Hc2kittaka_graph}. On the other hand, the angle dependence in the 2D limit is explained by the Tinkham model.
\begin{eqnarray}
\left|\frac{H_{\mathrm{c2}}(\theta)\sin \theta}{H_{\mathrm{c2}||c}}\right|+\left(\frac{H_{\mathrm{c2}}(\theta)\cos \theta}{H_{\mathrm{c2}||a}}\right)^2=1
\end{eqnarray}
As shown in Figs. 2(c) and (d), angle dependence observed in the {\SRO} film is fitted better by the 2D model. Assuming that both {\Hctwoa} and {\Hctwoc} are determined by the orbital limit as described by the GL equations, the effective superconducting thickness $d$ is estimated at 23 nm from $d=\sqrt{6\Phi_{0}H_{\mathrm{c2}||c}/\pi {H_{\mathrm{c2}||a}}^2}$. Considering that the film thickness is 50 nm, the film can be understood to be located in a dimensional crossover region.

In a very low angle region, $H_{\mathrm{c2}}$ seems suppressed compared to the 2D model. One possible origin of the deviation is the 2D-3D crossover. In such an intermediate superconducting state, the following empirical model interpolating Eqs. 1 and 2 has been proposed to explain the transitional angle dependence \cite{quasi2D1}.
\begin{eqnarray}
\alpha \left|\frac{H_{\mathrm{c2}}(\theta)\sin \theta}{H_{\mathrm{c2}||c}}\right|+(1-\alpha)\left(\frac{H_{\mathrm{c2}}(\theta)\sin \theta}{H_{\mathrm{c2}||c}}\right)^2 \nonumber \\
+\left(\frac{H_{\mathrm{c2}}(\theta)\cos \theta}{H_{\mathrm{c2}||a}}\right)^2=1
\end{eqnarray}
The curve fitting is improved by adopting this model with $\alpha$ ranging about from 0.8 to 0.9 (for details see Supplemental Material \cite{supplemental}), also suggesting that the system is located in the crossover region. {\Hctwo} around {\Hparaa} may be affected also by the presence of the Pauli limit, as discussed later.

Fig. 3(a) summarizes the $H$-$T$ phase diagram obtained for the {\SRO} film. Surprisingly, {\Hctwo} for {\Hparac} shows linear temperature dependence down to the lowest temperature without suppression as in the Werthamer-Helfand-Hohenberg (WHH) theory \cite{WHH1}, as also clearly confirmed in the raw data in Fig. 1(f). The linear dependence without any suppression at low temperatures may be related to $d$-vector flipping from perpendicular to parallel to the $ab$ plane in thin films, where the Pauli limit is no longer effective for the out-of-plane direction. For {\Hparaa}, {\Hctwo} follows the WHH-like curve but is rather weakly suppressed at low temperatures, in comparison to the bulk, as clearly seen in Fig. 3(b). Such a deviation from clear square-root temperature dependence expected in the 2D GL model has been also confirmed in other crossover systems showing the transitional field angle dependence with $0< \alpha <1$ \cite{quasi2D2}. $h^{\ast}$, {\Hctwo} normalized by the initial slope at {\Tc}, is saturated at about 0.64 for {\Hparaa}, which is even higher than the value of 0.42 measured for the bulk \cite{Hc2kittaka_graph}.

While the superconducting state approaches 2D like in the {\SRO} thin film, the anisotropy ratio $\Gamma=H_{\mathrm{c}2\parallel a}/H_{\mathrm{c}2\parallel c}$ itself is reduced to 10 near {\Tc} and 6 at the lowest temperature. This primarily results from increase in {\Hctwoc}, about four times over the bulk. As confirmed in Fig. 2(a), {\Hctwo} is anomalously enhanced over a wide range of field angles centered at {\Hparac}. Figure 4(a) plots the correlation between {\Hctwoc} and {\Tc} for {\SRO} bulks and films including previously reported other superconducting samples \cite{YoshiharuPLD, UchidaMBE}. Almost independent of the sample quality, the bulk and film {\Hctwoc} follow each universal curve, which is  roughly proportional to $T_{\mathrm{c}}^2$ as expected for the orbital-limiting {\Hctwo}. In the case of dirty samples, $\xi$ decreases with decrease of the mean free path $l$. This results in the extrinsic enhancement of {\Hctwo}, and this trend can be confirmed for MgB$_2$ and YBa$_2$Cu$_3$O$_7$ as positive correlation in the $l$-$\xi$ plot in Fig. 4(b). In the case of clean samples, on the other hand, $\xi$ increases with decrease of $l$, accompanied by the decrease of {\Tc} or superconducting gap $\Delta_{0}$. This trend appears as negative correlation in the $l$-$\xi$ plot. The {\SRO} films and also bulks independently show the clean-limit trend, excluding the extrinsic effects as a possible origin of increase of {\Hctwoc}.

By assuming that the GL in-plane coherence length $\xi_{ab}=\sqrt{\Phi_{0}/2\pi H_{\mathrm{c2}||c}}$ is equal to the Pippard one $\xi_{ab, 0}=\hbar v_{\mathrm{F}, ab}/\pi\Delta_0$ at the lowest temperature and using the superconducting gap relation $2\Delta_{0}=ak_{\mathrm{B}}T_{\mathrm{c}}$, the following relation can be derived.
\begin{eqnarray}
\frac{H_{\mathrm{c}2\parallel c}}{{T_{\mathrm{c}}}^2}=\frac{\pi\Phi_{0}}{8}\left(\frac{ak_{\mathrm{B}}}{\hbar v_{\mathrm{F}, ab}}\right)^2
\end{eqnarray}
In the right hand side, material dependent parameters are only the coupling ratio $a$ and the in-plane Fermi velocity $v_{\mathrm{F}, ab}$. For example, if we assume the BCS limit $a=3.5$ and take an experimental value of $v_{\mathrm{F}, ab}=9.3 \times 10^4$ m/s averaged on the active $\gamma$ band \cite{FSparameterization, BROstrain}, the dashed curve in Fig. 4(a) is obtained in rough agreement with but somewhat below the bulk trend, although a detailed analysis is surely dependent on the momentum-dependent gap structure \cite{Hc2deguchigap2} as well as the multi-band effect \cite{Hc2deguchigap1phase2}. In the case of thin films, on the other hand, other intrinsic origins should cause the further enhancement of {\Hctwoc} from the bulk trend. In terms of the epitaxial strain effect, a change in the in-plane lattice parameter compared to bulks is as small as $-0.07$\% at room temperature \cite{UchidaMBE}, which can be further reduced to $+0.03$\% at low temperatures \cite{SROparameter, LSATparameter}. In addition, angle-resolved photoemission spectroscopy on strained {\SRO} films grown on various substrates has demonstrated that the in-plane effective mass shows weak monotonic dependence on the strain value (less than 5\% for the 1\% in-plane lattice change) for all the three bands \cite{BROstrain}, indicating that $v_{\mathrm{F}, ab}$ is not a principal factor determining the enhancement. The uniaxial strain effect on bulks and films \cite{Hc2strain, HariMBE} is also excluded, as the present tiny strain is biaxial. Instead, an increase in the coupling ratio $a$ (almost double) is one plausible origin. The enhancement of {\Hctwoc} and the rather two-dimensional-like field-angle dependence are commonly observed in the films, regardless of the definition of {\Tc} nor the film quality, as shown in Supplemental Material \cite{supplemental}. Because all the other possible origins, such as film quality, epitaxial strain, and quantum confinement, are carefully excluded, the enhancement of {\Hctwoc} or $a$ is most likely related to the observed dimensional crossover. Electrons may couple more strongly in the real space through the dimensional crossover, resulting in the shorter $\xi_{ab}$. Its microscopic mechanism will need to be further elucidated from theoretical aspects, while it may be also consistent with the recent theoretical prediction on two-dimensional {\SRO} films, as discussed below.

While {\Hctwo} is largely enhanced centered at {\Hparac}, it remains relatively low for {\Hparaa}. One origin of this difference is a change in the out-of-plane electronic structure by quantum confinement in films. However, an increase in the out-of-plane Fermi velocity $v_{\mathrm{F}, c}$, which may account for the elongation of $\xi_{c}$, is less likely in terms of the mass enhancement due to the confinement. Another possible origin of this relative suppression is the Pauli limit. The presence of the Pauli limit for {\Hparaa} is not generally consistent with the $d$-vector direction ($d \parallel c$) in the 2D {\pxy} state \cite{SROreview4desiringfilm}. Therefore, the suppression of {\Hctwoa} suggests a change of the pairing symmetry, possibly accompanied with the $d$-vector flipping ($d \parallel ab$) suggested for thin films \cite{thinSROdvector}. This is also consistent with the disappearance of the suppression newly observed in the temperature dependence of {\Hctwoc} (Fig. 3(b)), indicating the absence of the Pauli limit for {\Hparac} in thin films.

In summary, we have revealed changes of the {\SRO} superconducting state induced by confining it into thin films. Through the dimensional crossover, {\Hctwo} is intrinsically enhanced centered at {\Hparac} compared to bulks, while it remains suppressed for {\Hparaa}. The anomalous enhancement of {\Hctwo} suggests important changes of the spin-triplet superconducting state in the reduced dimensions. Taken together, these findings are compatible with the triplet state with the $d$-vector flipped parallel to the RuO$_2$ plane, which still could support the Majorana modes at edges and vortices \cite{thinSROdvector}. Our study will provide the significant basis for further investigating superconducting properties of {\SRO} thin films and applying its exotic states to junction devices.

We acknowledge T. Nojima, S. Kittaka, D. Manske, T. Shibauchi, Y. Mizukami, Y. Iwasa, Y. Saito, and Y. Nakagawa for fruitful discussions. This work was supported by Grant-in-Aids for Scientific Research (B) No. JP18H01866 and Scientific Research on Innovative Areas ``Topological Materials Science" No. JP16H00980 from MEXT, Japan and JST PRESTO No. JPMJPR18L2 and CREST Grant No. JPMJCR16F1, Japan.

\clearpage

\begin{figure}
\begin{center}
\includegraphics*[width=14.5cm]{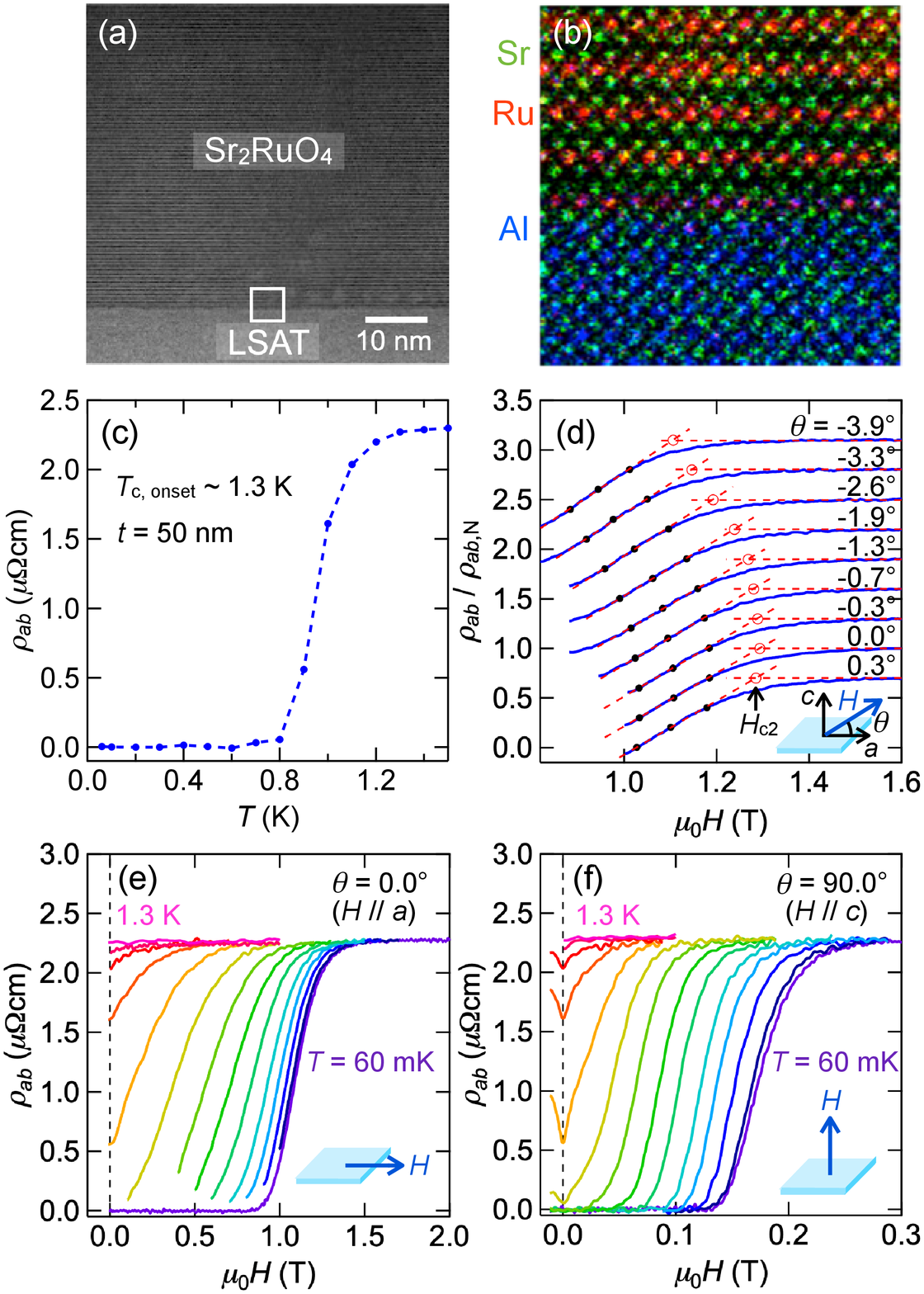}
\caption{
Characterization of a superconducting {\SRO} thin film. (a) Cross-sectional transmission electron microscope image and (b) its magnification in the boxed area, colored by energy dispersive x-ray spectrometry for Sr $K$, Ru $L$, and Al $K$ edges. (c) Temperature dependence of the in-plane resistivity $\rho_{ab}$, taken for the {\SRO} film with the transition temperature of $T_{\mathrm{c}}\sim1.3$ K (onset) and the film thickness of $t=50$ nm. (d) Field dependence of $\rho_{ab}$ measured around $\theta=0^{\circ}$ at $T=0.1$ K. Here, $\theta$ denotes the angle between the magnetic field and the $a$ axis within the $ac$ plane. The data are normalized by the normal-state in-plane resistivity $\rho_{ab, \mathrm{N}}$. An open circle represents {\Hctwo} defined as the intersection between two dashed lines extrapolated from normal ($\rho_{ab, \mathrm{N}}$) and superconducting (0.3--0.7$\rho_{ab, \mathrm{N}}$) regions. The points with resistivity of 0.3$\rho_{ab, \mathrm{N}}$, 0.5$\rho_{ab, \mathrm{N}}$, and 0.7$\rho_{ab, \mathrm{N}}$ are denoted by a filled circle. (e) and (f) In-plane ($\theta=0^{\circ}$) and out-of-plane ($\theta=90^{\circ}$) field dependence of $\rho_{ab}$ at the lowest temperature of $T=60$ mK and from 0.1 to 1.3 K at intervals of 0.1 K.
}
\label{fig1}
\end{center}
\end{figure}
\newpage

\begin{figure}
\begin{center}
\includegraphics*[width=14.5cm]{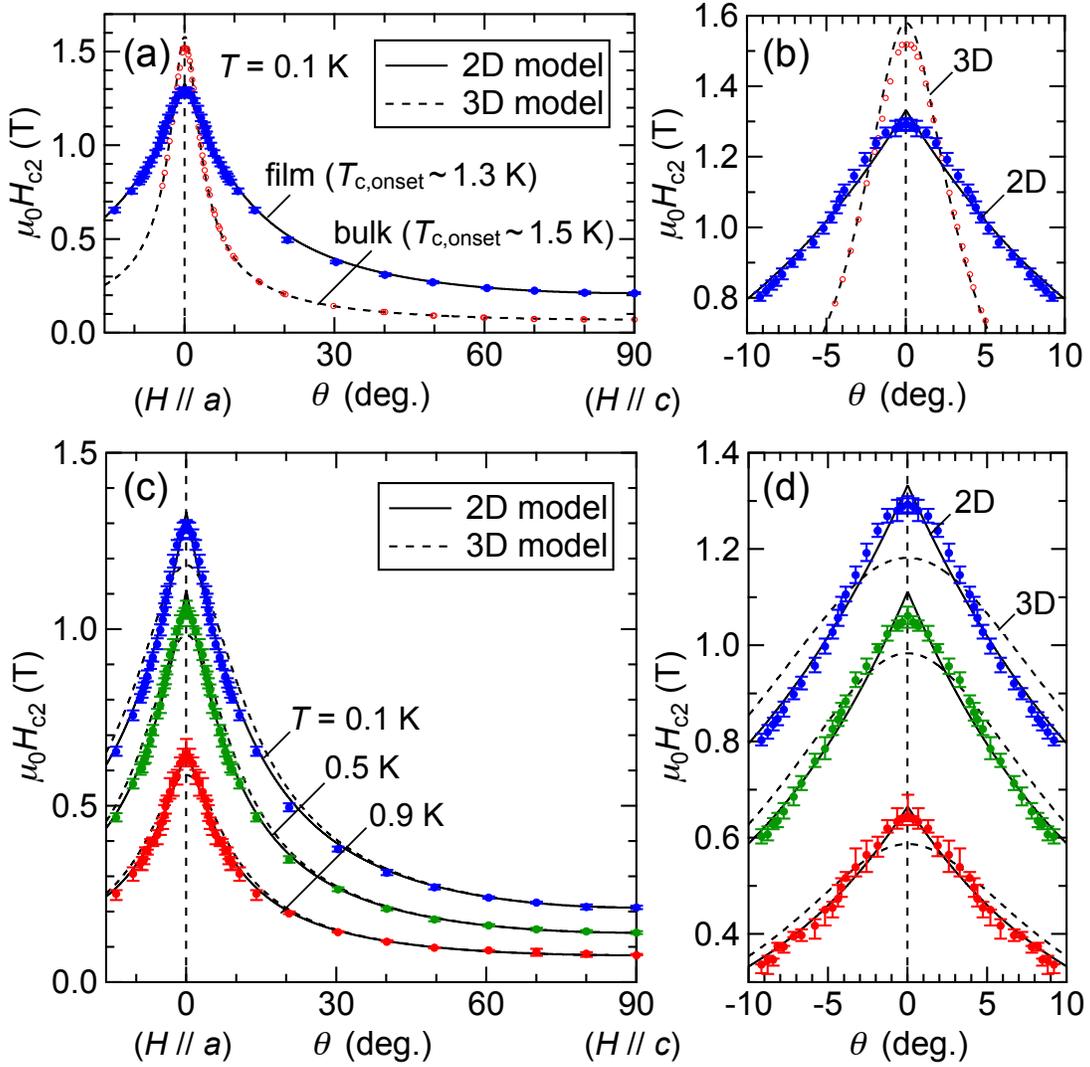}
\caption{
Dimensional crossover of the {\SRO} superconducting state. (a) Field angle dependence of {\Hctwo} in the {\SRO} film at $T=0.1$ K, compared to bulk one previously reported in Ref. \cite{Hc2kittaka_graph}. Dashed and solid curves are fitting results using the three-dimensional (3D) Ginzburg-Landau (GL) anisotropic mass model (Eq. 1) and the two-dimensional (2D) Tinkham model (Eq. 2), respectively. An enlarged view centered at $\theta=0^{\circ}$ is shown in (b). (c) and (d) Field angle dependence in the film at different temperatures fitted by the 2D and 3D models and its magnification around $\theta=0^{\circ}$. The field angle dependence in the film is described better by the 2D model.
}
\label{fig2}
\end{center}
\end{figure}
\clearpage

\begin{figure}
\begin{center}
\includegraphics*[width=14.5cm]{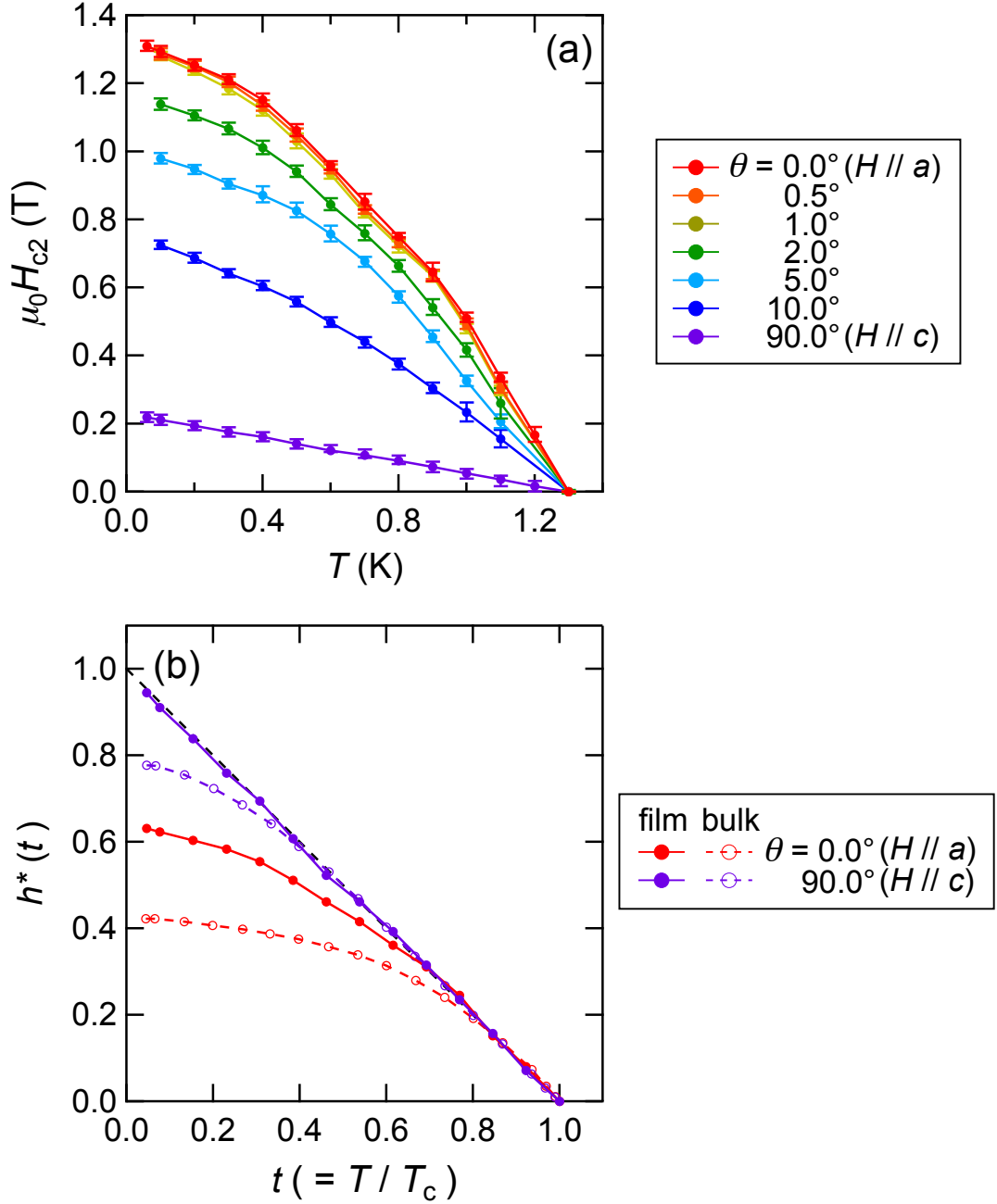}
\caption{
Superconducting phase diagram in the crossover region. (a) $H$-$T$ phase diagram of superconductivity in the {\SRO} film at various field angles between $\theta=0^{\circ}$ and $90^{\circ}$. (b) Temperature dependence of the normalized upper critical field $h^{\ast}(t)$ defined as $h^{\ast}(t)=-H_{\mathrm{c2}}(t)/(\mathrm{d}H_{\mathrm{c2}}/\mathrm{d}t)|_{t=1}$ ($t=T/T_{\mathrm{c}}$), compared to the bulk one \cite{Hc2kittaka_graph}.}
\label{fig3}
\end{center}
\end{figure}
\clearpage

\begin{figure}
\begin{center}
\includegraphics*[width=9.5cm]{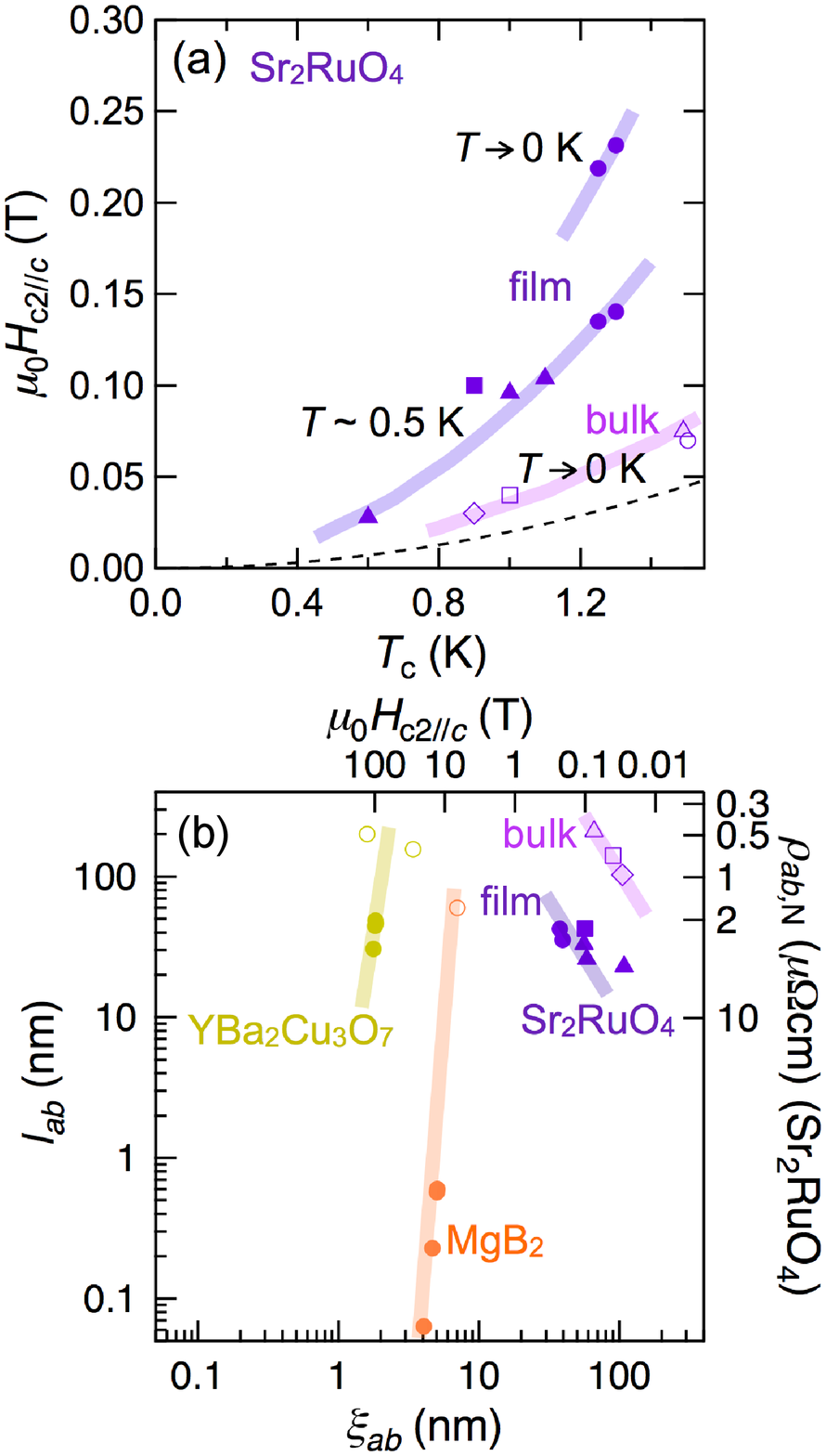}
\caption{
Enhancement of upper critical field in thin films. (a) {\Hctwoc} plotted as a function of {\Tc}, including data previously reported for superconducting {\SRO} bulks and films. As represented by the zero-temperature values deduced in this study, {\Hctwoc} are systematically enhanced in thin films ($\CIRCLE$ (present study), \ding{115} \cite{UchidaMBE}, $\blacksquare$ \cite{YoshiharuPLD}), in comparison to bulk ones ($\Circle$ \cite{Hc2kittaka_graph}, $\bigtriangleup$ \cite{Hc2akimacohelength_graph}, $\square$ \cite{Hc2mackenzie_graph}, $\Diamond$ \cite{Hc2yoshida_graph}). The dashed curve is calculated following Eq. 4. (b) Mean free path $l_{ab}$ vs coherence length $\xi_{ab}$ summarized for the {\SRO} bulks and films. $l_{ab}$ is estimated from the common-$l$ approximation $l_{ab}=hc/2e^2 \rho_{ab, \mathrm{N}} \sum_{i} k_{\mathrm{F},i}$, with the interlayer spacing $c/2$ and the $i$-th Fermi wave number $k_{\mathrm{F},i}$ \cite{Tcmackenzie, Hc2akimacohelength_graph}. The corresponding $\rho_{ab, \mathrm{N}}$ and $H_{\mathrm{c2}||c}$ are labeled on the right and top axes, respectively. For reference, data in MgB$_2$ \cite{Hc2MgB2} and YBa$_2$Cu$_3$O$_7$ \cite{Hc2YBCO1bulk, Hc2YBCO2film} are also presented for bulks and films as denoted by open and closed symbols.
}
\label{fig4}
\end{center}
\end{figure}
\clearpage

\end{document}